\documentclass[manuscript]{aastex}

\usepackage{graphicx}
\usepackage{multirow}
\usepackage{xcolor}
\usepackage{lscape}
\usepackage{hyperref}
\usepackage[normalem]{ulem}

\shorttitle{Photometric analysis of four recently discovered contact binaries}
\shortauthors{Djura{\v s}evi{\' c} et al.}

\begin{document}

\title{A photometric study of four recently discovered contact binaries: \\
	1SWASP J064501.21+342154.9,	1SWASP J155822.10-025604.8, \\
	1SWASP J212808.86+151622.0 and UCAC4 436-062932}

\author{G. Djura{\v s}evi{\' c}}
\affil{Astronomical Observatory, Volgina 7, 11060 Belgrade, Serbia}
\affil{Isaac Newton Institute of Chile, Yugoslavia Branch}
\email{gdjurasevic@aob.rs}

\author{A. Essam}
\affil{Department of Astronomy, National Research Institute of Astronomy and Geophysics, Helwan, Egypt}
\email{essam60@yahoo.com}

\author{O. Latkovi{\' c}}
\affil{Astronomical Observatory, Volgina 7, 11060 Belgrade, Serbia}

\author{A. Cs{\'e}ki}
\affil{Astronomical Observatory, Volgina 7, 11060 Belgrade, Serbia}

\author{M. A. El-Sadek}
\affil{Department of Astronomy, National Research Institute of Astronomy and Geophysics, Helwan, Egypt}

\author{M. S. Abo-Elala}
\affil{Department of Astronomy, National Research Institute of Astronomy and Geophysics, Helwan, Egypt}

\and

\author{Z. M. Hayman}
\affil{Department of Astronomy and Numerology and Space Science, Faculty of Science, Cairo University, Cairo, Egypt}

\begin{abstract}

We present new, high quality multicolor observations of four recently discovered contact binaries: \object{1SWASP J064501.21+342154.9}, \object{1SWASP J155822.10-025604.8}, \object{1SWASP J212808.86+151622.0} and \object{UCAC4 436-062932}, and analyze their light curves to determine orbital and physical parameters using the modeling program of G. Djura{\v s}evi{\' c}. In the absence of spectroscopic observations, the effective temperatures of the brighter components are estimated from the color indices, and the mass ratios are determined with the q-search method. The analysis shows that all four systems are W~UMa type binaries in shallow contact configurations, consisting of late-type main sequence primaries and evolved secondaries with active surface regions (dark or bright spots) resulting from magnetic activity or ongoing transfer of thermal energy between the components. We compare the derived orbital and stellar parameters for these four variables with a large sample of previously analyzed W~UMa stars and find that our results fit it well.

\end{abstract}


\section{Introduction}

Low-temperature, short-period contact binaries, also known as W~UMa stars, are among the most extreme and least understood stages of binary evolution and interaction. A remarkable feature of these stars is that, even for small mass ratios (large differences between masses), the temperatures of components are typically close to equal (with differences in the order of 100 K), which is understood in terms of efficient heat transfer through a common convective envelope \citep{lucy1968a, lucy1968b}. Components of a close binary system may come in contact when the more massive (and thus more quickly evolving) star fills its Roche lobe during its normal post-main-sequence expansion, and deposits a large amount of material on its companion, which then in turn fills its own Roche lobe, so that a common envelope is formed. This and alternative evolutionary paths from detached to contact binaries are discussed in detail by \citet{yakut2005} and \citet{eggleton2006}.

Light curves of W~UMa stars are distinguished by continuous changes in brightness resulting from ellipsoidal variations, minima of nearly equal depths, and maxima that aren't always symmetric. This difference in maximum light levels, sometimes referred to as the O'Connell effect \citep{oconnell1951}, is caused by inhomogeneity in the surface brightness distribution on one or both stars and is commonly associated with dark spots of magnetic origin.

W~UMa stars are traditionally divided in two groups: the A and W types \citep{binnen}. In A type systems, the more massive component is eclipsed in the primary (deeper) minimum; and in W type systems, the less massive. In addition to that, A types have smaller mass ratios and spectral types ranging from A to G, while W types have larger mass ratios and spectral types from F to K. There's a long-standing debate in the literature on whether these two groups represent an evolutionary sequence \citep[see e.g.][]{gazeas2006}, but \citet{yildiz2013} recently showed that the initial parameters from which the two types evolved are completely different. \citet{lucy1979} introduced the additional B type, or contact binaries in poor thermal contact, for variables of observational characteristics similar to other W~UMa stars, but where the temperatures of components differ by more than 1000 K.

W~UMa binaries make especially attractive subjects for photometric studies. With orbital periods typically shorter than 0.7 days \citep{hilditch2001}, several nights of observations are needed to obtain a complete light curve. For systems exhibiting high inclination, the mass ratios (necessary for a reliable orbital solution) can be inferred from purely geometric arguments even in the absence of complementary spectroscopic data \citep{terrell2005}. Suitable modeling techniques can then be applied to the light curves in order to estimate the absolute stellar parameters for the individual components (see e.g. \citealt{wilson1994} for a review of modeling methods).

In this study we analyze the light curves of four newly discovered contact systems: \object{1SWASP J064501.21+342154.9} (hereafter J0645), \object{1SWASP J155822.10-025604.8} (hereafter J1558), \object{1SWASP J212808.86+151622.0} (hereafter J2128) and \object{UCAC4 436-062932} (hereafter UCAC4) with the aim to determine their orbital and stellar characteristics. For this purpose we obtained high quality multicolor CCD observations from the 1.88 m telescope in Kottamia Observatory in Egypt, and constructed binary star models that optimally fit these observations using the modeling program of G. Djura{\v s}evi{\' c} \citep{djur92a, djur98}. With the  exception of J0645 (which has been analyzed recently by \citealt{liu14}), the present work is the first in-depth study of these variables.

Our findings suggest that all four systems are contact binaries of W~UMa type, with J0645, J1558 and UCAC4 belonging to the A type, and J2128 to the W type. In all four cases we use dark or bright spots on one or both components to model the asymmetries of the light curves and improve the fit of the models to observations. The dark spots are interpreted in terms of surface magnetic activity, while the spots in the neck region (which may be bright or dark) represent temperature inhomogeneities arising from the ongoing exchange of thermal energy between the stars through the common envelope.

Below we describe the details of the observations (Section ~\ref{observations}), light curve analysis (Section~\ref{analysis}) and discuss each system in turn (Sections~\ref{sJ0645},~\ref{sJ1558},~\ref{sJ2128} and~\ref{sUCAC4}).

\section{Observations}\label{observations}

Photometric observations of selected variables were carried out with the wide Bessell $V$, $R$ and $I$ passbands closely matched to the classic Johnson-Cousins system, using the back-illuminated EEV CCD 42-40 camera with 2048 $\times$ 2048 pixels. The pixel size, scale and total field of view are 13.5$\mu$, 0.305" per pixel, and 10 $\times$ 10 arcmin$^2$, respectively. The camera is attached to the Newtonian focus (f/4.84) of the 1.88 m reflector telescope of Kottamia Astronomical Observatory (KAO) in Egypt, and kept at a temperature of -125\degr\ by liquid nitrogen cooling. For more details about KAO instruments, see \citet{azzam}.

Exposure times for J0645 were 60, 20 and 10 s for the $V$, $R$ and $I$ passbands, respectively; 150, 80 and 60 s for J1558 and UCAC4; and 100, 40 and 20 s for J2128. All frames were reduced (bias-subtracted and flat-field corrected), and differential aperture photometry was performed using the software package C-Munipack\footnote{\url{ http://c-munipack.sourceforge.net/ }}. Finally the magnitudes (variable minus comparison) and their errors were computed in each passband.

Information about the variable, comparison and check stars was obtained from the UCAC4 Catalog \citep{zach12} and the NOMAD Catalog \citep{zach05}, and is listed in Table~\ref{tObsLog}. The comparison and check stars were chosen so as to match as closely as possible the position, magnitude and color of the corresponding program star.

Times of minima were calculated using the software package AVE \citep{barbera} which is based on the method of \citet{kwandvw56}, and are given in Table~\ref{tMin}. A preview of the light curve data is given in Table~\ref{tData}, and the full light curves are available as machine-readable tables.

\section{Light curve analysis}\label{analysis}

The analysis of the light curves was done using the program by \citet{djur92a} generalized for the case of contact configurations \citep{djur98}, which implements a sophisticated and versatile binary star model based on Roche geometry. This program has been used and tested for more than two decades on a wide range of binary configurations \citep[see e.g.][]{djur2010,djur2011,mennickent2015}. Its distinguishing features are the efficient approach to visibility detection during the eclipses, the treatment of the reflection effect, and the model optimization method. Namely, the parameters of the model are estimated using the Marquart-Levenberg algorithm \citep{marquardt} with modifications described in detail in \citet{djur92b} to minimize the sum of squared residuals between the observed ($O$) and calculated ($C$) light curves simultaneously in all passbands. 

A comprehensive list of all model parameters can be found below. Note that we consistently refer to the more massive star as the primary component (with the subscript 1), so that mass ratio, $q$, is always less than 1. Unless stated otherwise, the reflection coefficients (albedos) and gravity darkening exponents were kept fixed to their expected values for stars with convective envelopes \citep{lucy,vonz}. The effective temperature, corresponding to the average of local temperatures weighted by the areas of elementary surfaces, is estimated from the $B-V$ color index (see Table~\ref{tObsLog}) for the brighter component and adjusted as a free parameter for the other. Treatment of limb darkening follows the nonlinear approximation of \citet{claret00}, with the coefficients for the appropriate passbands interpolated from their tables based on the current values of effective temperature and effective gravity in each iteration. The reflection effect is accounted for by applying a temperature correction to affected elementary surfaces, which is calculated following the prescription of \citet{khruzina1985}.

\renewcommand{\labelitemi}{$\circ$}
\begin{footnotesize}
\begin{itemize}
\itemsep0pt
\item Point count --- the total number of the observations spanning all the  passbands.
\item $\rm \sum{(O-C)^2}$ --- the final sum of squares of residuals between the observed (LCO) and synthetic (LCC) light-curves (in magnitudes).
\item P --- the orbital period, in days.
\item $\rm q=m_2/m_1$ --- the mass ratio of the components.
\item i --- the orbital inclination (in degrees).
\item $\rm a_{ORB}$ --- the orbital semi-major axis in units of solar radius.
\item $\rm \ell_3/(\ell_1+\ell_2+\ell_3)$ --- the contribution of uneclipsed (third) light to the total light of the system at the phase of the light-curve maximum (omitted when zero).
\item $\rm f_{over}$ --- the degree of overcontact, or overflow (in percents), defined as $f_{over}=100 \frac{\Omega - \Omega_{in}}{\Omega_{out} - \Omega_{in}}$.
\item $\rm \Omega_{in}\ and\ \Omega_{out}$ --- the dimensionless values of Roche potential at the inner and outer critical surfaces that contain the equilibrium points $\rm L_1$ and $\rm L_2$, respectively.
\item A, $\beta$ --- the albedo and the gravity-darkening exponent of the component.
\item $\rm T_{eff}$ --- the effective temperature of the component (in Kelvins).
\item F --- the filling factor of the component, defined as the ratio between the stellar polar radius and the polar radius of the critical Roche surface. In contact systems, this quantity is the same for both stars and we treat it as a system parameter.
\item $\rm \Omega$ --- the dimensionless surface potential of the component.
\item $\rm L/(L_1+L_2)$ --- the contribution of the component to the total luminosity of the system.
\item R --- the polar radius of the component in units of separation.
\item $\cal M, \ \cal R$ --- the mass and the mean radius of the component in solar units.
\item $\rm \log g$ --- the logarithm (to base ten) of the effective gravity of the component in CGS units.
\item $\rm M_{bol}$ --- the absolute bolometric magnitude of the component.
\item $\rm T_{spot}/T$ --- the ratio between the temperature of the spot and the local temperature of the star.
\item $\rm \theta,\ \lambda,\ \varphi$ --- the angular radius, longitude, and latitude of the spot (in arc degrees).
\end{itemize}
\end{footnotesize}

\begin{figure*}
\includegraphics{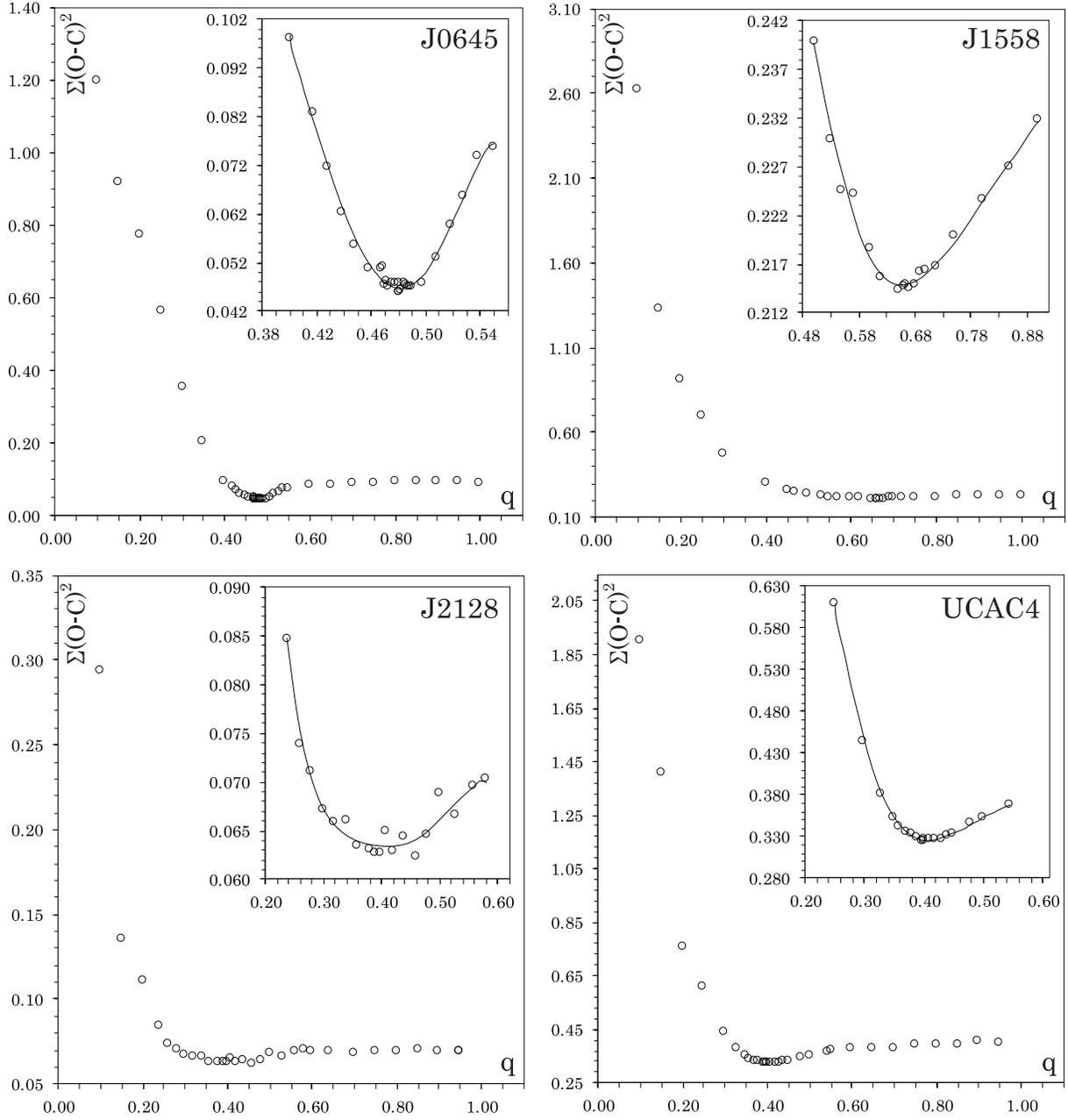}
\caption{The \textit{q-search} results. Each point represents the quality of fit (the sum of squared $O-C$ residuals) for the candidate mass ratio. The insets zoom in on the regions around the best solution, and show the polynomial fit (solid line) used to find the minimum.}
\label{fQS}
\end{figure*}

Of particular interest is the determination of the mass ratios. In the absence of spectroscopic observations and radial velocity studies for these recently discovered variables, the mass ratios were estimated from photometric data only, using the so-called q-search method (illustrated in Figure~\ref{fQS}). This approach entails making a sequence of simplified, preliminary models (that, for example, do not include spots or the third light contribution) with different values of mass ratio, selected so as to uniformly cover a reasonable range; in this study, the q-search is performed from $q=0.1$ to $1.0$. These models are then optimized to fit the observations as well as possible by adjusting the major orbital and stellar parameters (inclination, filling factors, temperatures) - but not the mass ratio itself, which is kept fixed at its initial value. The quality of the fit, quantified with the sum of squared residuals between the observed and calculated light curves, $\Sigma(O-C)^2$, is then plotted against the mass ratio for each trial value of $q$. Since the mass ratio affects the relative sizes of the components, it has an appreciable effect on the shape of the light curve, especially in systems with inclinations high enough to produce total eclipses. For this reason we expect that the quality of fit will significantly increase as the trial values of $q$ approach the actual mass ratio. The minimum of the q-search curve, representing the best mass ratio, is found by fitting a low-order polynomial through the points (insets on Figure~\ref{fQS}). Then the procedure is repeated for a finer sampling of trial values of $q$ in a region near this minimum. The model with the best candidate mass ratio is finally optimized with the full set of parameters, $q$ included.

\begin{figure*}
\includegraphics{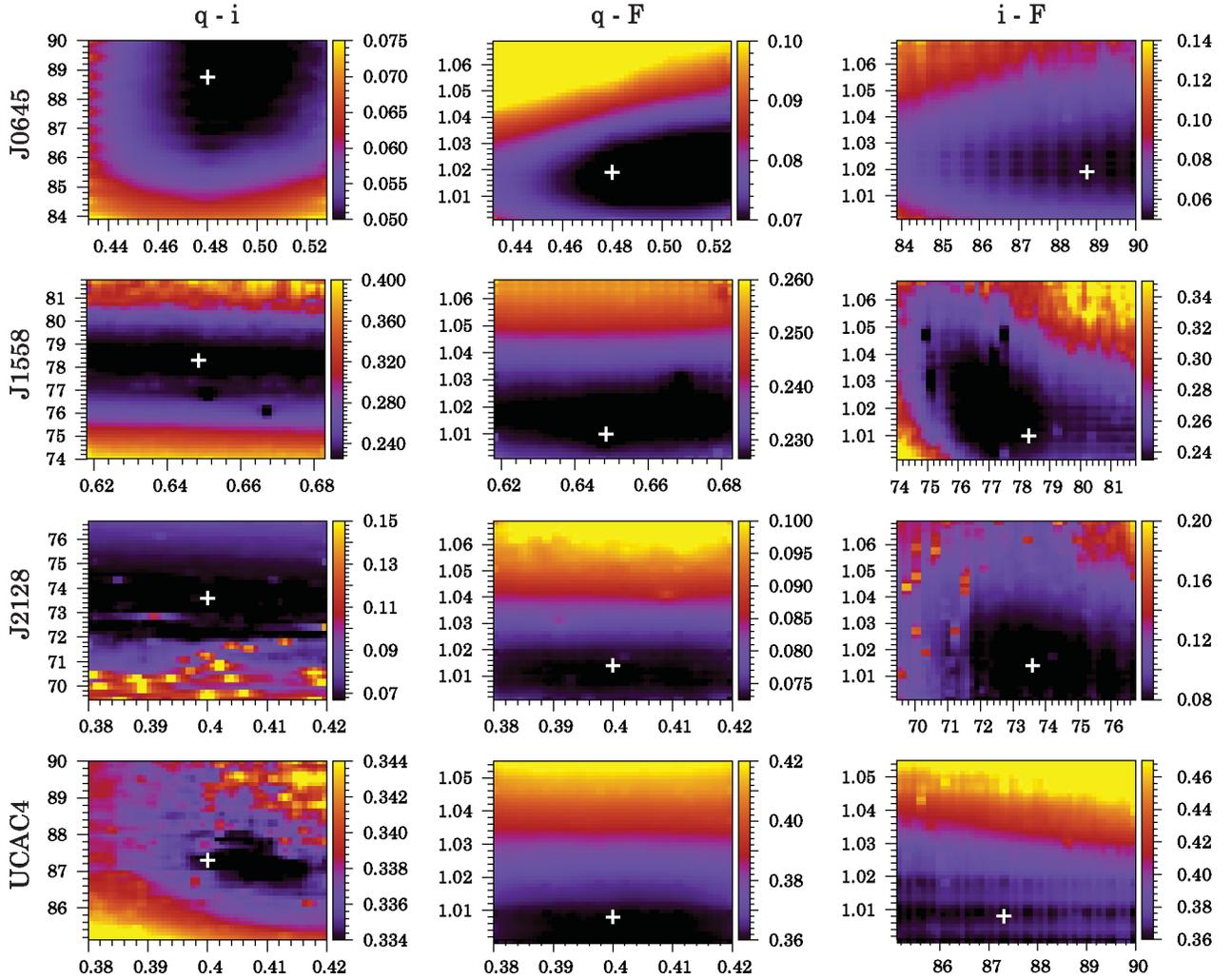}
\caption{Parameter space mapping in the vicinity of optimal models. Each row corresponds to a star, and each column to a correlation between: mass ratio and inclination (left), mass ratio and filling factor (middle) and inclination and filling factor (right). The quality of fit of synthetic to observed light curves, $\Sigma(O-C)^2$, is color-coded with darker colors corresponding to smaller values (better fits), and the optimal solutions given in Tables~\ref{tJ0645},~\ref{tJ1558},~\ref{tJ2128} and~\ref{tUCAC4} are represented with white crosses.}
\label{fStab}
\end{figure*}

The stability of these solutions was tested with a heuristic scan of the surrounding parameter space. We chose three pairs of highly correlated parameters ($q-i$, $q-F$ and $i-F$) and made two-dimensional grids of perturbed models with 50 values along each dimension (2500 models per grid) in the range of $\pm 5\%$ from the values in the final solution found by the q-search. The other key parameters (primary or secondary temperature and the sizes, locations and contrasts of the spots) were perturbed at random in the same $\pm 5\%$ range\footnote{Some parameters have natural limits (i.e. the inclination cannot be greater than $90\degr$ and the filling factor cannot be less than 1 in contact configuration). In such cases the range of perturbations is smaller and asymmetric.}. All these models were then optimized for a best fit to the observations, and the goodness of achieved fit can be used to examine the parameter space. This is illustrated in Figure~\ref{fStab}, which shows the $q-i$, $q-F$ and $i-F$ model grids as contour plots where the quality of fit is color-coded so that darker colors correspond to smaller values (better fit) and vice versa. Finally, we picked the best parameter combinations from these plots (the ones with smallest $\Sigma(O-C)^2$, shown in Figure~\ref{fStab} as white crosses), and used them as initial values for one last model optimization. It is this final model that is presented in subsequent sections on individual stars.

To derive the absolute system parameters, the mass of the brighter component was estimated based on the spectral type - color index calibration from \citet{lang}, assuming the star is on the main sequence. The mass of the other component can then be calculated from the mass ratio. Knowing the total mass and the period of the system, we derive the the orbital separation from third Kepler's law.

The uncertainties in absolute orbital and stellar parameters provided in Tables \ref{tJ0645}, \ref{tJ1558}, \ref{tJ2128}, \ref{tUCAC4} and throughout the text are derived from the formal fitting errors, assuming an error of one subtype in the determination of spectral type.

Now we present the results of our work for each system in turn.

\section{J0645}\label{sJ0645}

The variability of J0645 was discovered by \citet{norton}, who measured a period of 0.22105 days, a maximum of 14.11 mag and the depth of minima of 0.35 in the V band. Later \citet{lohr} determined that the orbital period is 0.2486159 days, and that is the value used in this study.

J0645 was recently observed and studied in more detail by \citet{liu14}, who found a photometric mass ratio of $q=m_2/m_1=0.474$, a degree of overcontact of $f_{over}\approx 15\%$, minor third light contribution in the I band and a small dark spot with a considerable temperature contrast ($T_{spot}/T_*=0.8$) on the larger, cooler and more massive star\footnote{Note that \citet{liu14} denote the larger, cooler and more massive star eclipsed in the deeper minimum with subscript 2, while we denote it with subscript 1.}.

We observed J0645 on February 7$^{th}$, 2013 and measured times of minimum light on three other nights (January 4$^{th}$ and 5$^{th}$, and February 8$^{th}$, 2013).  A preliminary analysis of these data was done by \citet{essam13}, who manually tweaked the major system parameters to get a visually good fit of the calculated light curves to the observations, with the aim of providing rough first estimates. Such procedure is hard to replicate, verify and reproduce, which is why we reanalyze the same data in this work, using the results of \citet{essam13} as the starting point for our automatic model optimization.

The light curves were folded using the following ephemeris:

\begin{equation}\label{ephJ0645}
Min_I [HJD] = 2456331.3940(2) + 0.2486159\times E
\end{equation}

The mass ratio of $q=m_2/m_1=0.48$, determined by the q-search method (Figure~\ref{fQS}, top left), can be considered reliable given the total eclipse, which constrains the orbital inclination to within a couple degrees from an edge-on orientation. This result is in remarkably good agreement with the value calculated by \citet{liu14}. In our solution, the system is in a contact configuration with $f_{over}\approx 16\%$ and has nearly identical temperatures of the components ($\Delta T\approx 130$K). 

Since the more massive star is also the one eclipsed in the deeper minimum, J0645 belongs to the A type of W~UMa binaries according to our solution. The late spectral types of the components and the low total mass suggest that it might belong to the W type instead, which is how \citet{liu14} classify it; but we could not find a solution as good as the one presented here for that configuration. \citet{liu14} assign the zero orbital phase (the "primary minimum") to the total eclipse, while we assign it to the partial, but slightly deeper eclipse, which accounts for the difference. Minima of virtually equal depth are no rarity among W~UMa stars, and in such cases only a radial velocity study can lift the ambiguity between the subtypes with certainty.

There is a noticeable asymmetry in the light curves before and after the primary minimum, which is well accounted for by including in the model a small dark spot with the temperature contrast of $T_{spot}/T_*=0.96$ on the primary star. Although the location of this spot differs by about $30\degr$ in longitude from the spot in the solution of \citet{liu14}, is of greater size and smaller temperature contrast, these results are qualitatively similar and point to surface activity of magnetic nature on the primary star.

The fit of the synthetic light curve to the observations can be slightly improved by allowing for the presence of uneclipsed (third) light and treating it as a free parameter of the model. The contribution of third light is nearly negligible, at $\rm \ell_3/(\ell_1+\ell_2+\ell_3)=0.0$ in the V band and $0.01$ in the R and I bands, however it confirms the results obtained by \citet{liu14}.

\begin{figure*}
\includegraphics{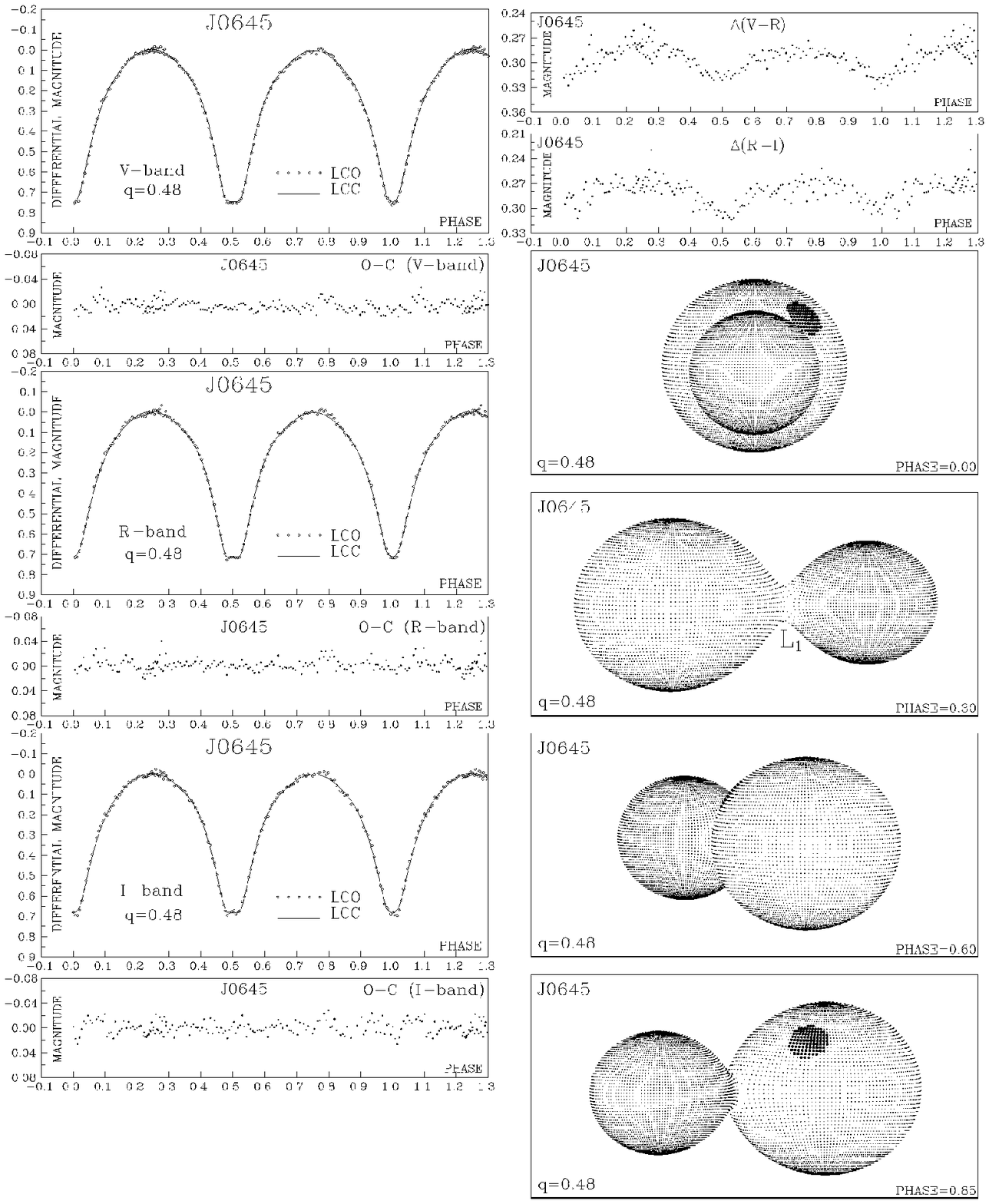}
\caption{Observed (LCO) and synthetic (LCC) light curves and the final $O-C$ residuals of {\rm J0645}, with $\Delta(V-R)$ and $\Delta(R-I)$ color curves and the graphic representation of the model described in Section~\ref{sJ0645} at the orbital phases 0.00, 0.30, 0.60 and 0.85.}
\label{fJ0645}
\end{figure*}

Our findings are summarized in Table~\ref{tJ0645} and Figure~\ref{fJ0645}, that shows the observed and simulated light curves with residuals, the color curves and the representation of the model at several orbital phases.

\section{J1558}\label{sJ1558}

J1558 was found to be a variable star by \citet{norton}, who reported a period of 0.23008 days, a maximum of 13.77 mag and the depth of minima of 0.18 in the V band. \citet{lohr} gives an improved period of 0.2600776 days, which is the value used in this study. To our knowledge there have been no other observations or studies of J1558 to date, making this the first quantitative analysis of the variable.

Our observations of J1558 were made on three nights (May 22$^{nd}$, and June 17$^{th}$ and 18$^{th}$, 2014). The following ephemeris was used to fold the light curve to orbital phases:

\begin{equation}\label{ephJ1558}
Min_I [HJD] = 2456826.3430(3) + 0.2600776 \times E
\end{equation}

As in the case of J0645, the analysis of the light curves of J1558 begins with the determination of the mass ratio. The q-search (Figure~\ref{fQS}, top right) gives the value of $q=m_2/m_1=0.65$. Assuming this mass ratio, the system is in contact configuration with $f_{over}\approx 7\%$ and the orbital inclination of $i\approx 78\degr$. With the more massive star eclipsed in the primary minimum, J1558 is A type W~UMa binary.

The temperature of the primary component was determined from the $B-V$ color index and kept fixed at $T_1=6200$ K, while the optimization of the model results with $T_2=5970$ K. The fit of the model to the observations could be significantly improved by allowing the albedos of the components to be adjusted as free parameters. This resulted in albedos higher than theoretical values for stars with convective envelopes, which is an indication of possible presence of a bright spot in the neck region. And indeed a model with such a spot (and albedo values fixed to their theoretical values) fits the observations remarkably well. We interpret this finding as a consequence of ongoing thermal exchange between the components through the neck region of the common envelope.

\begin{figure*}
\includegraphics{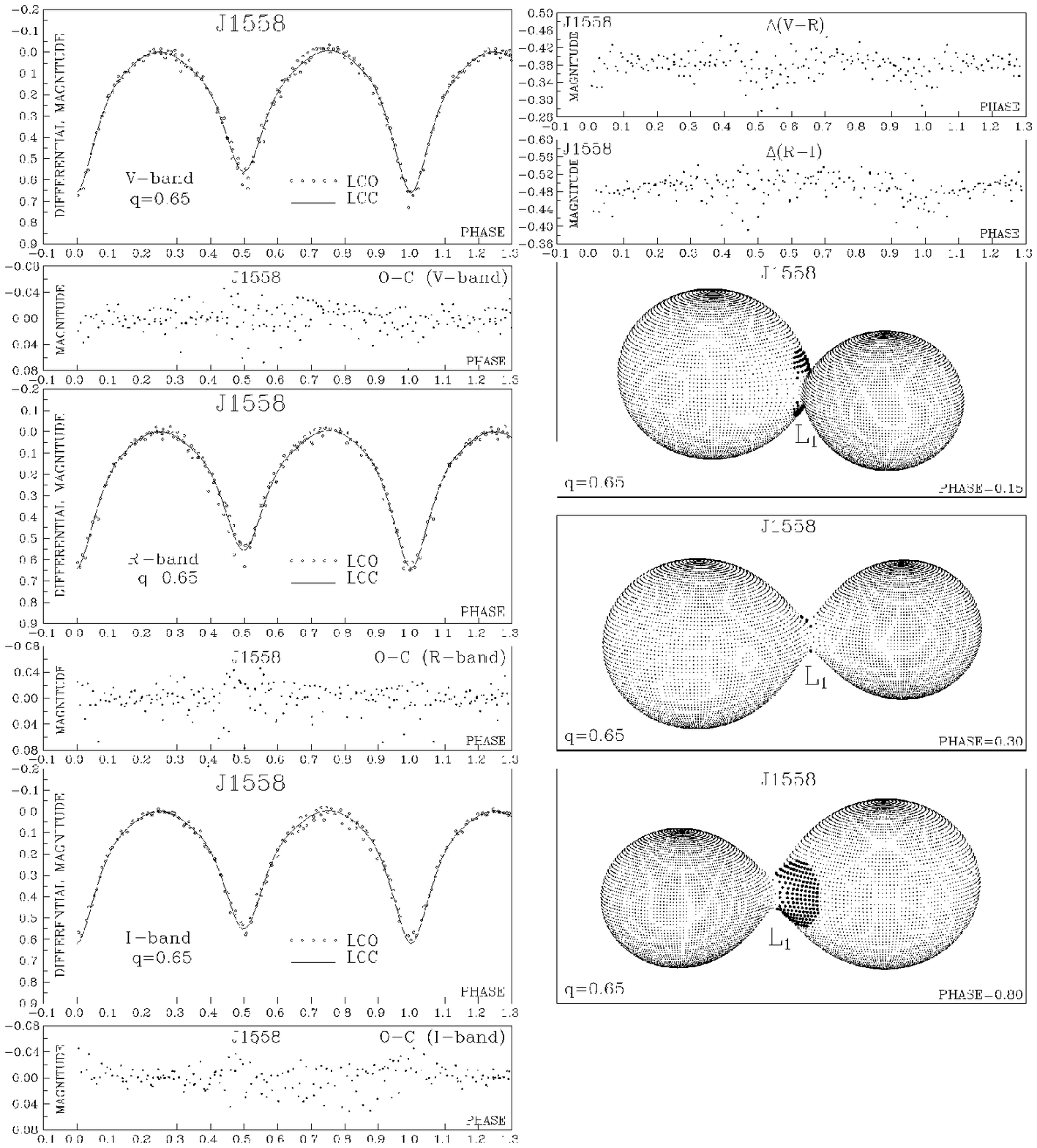}
\caption{Observed (LCO) and synthetic (LCC) light curves and the final $O-C$ residuals of {\rm J1558}, with $\Delta(V-R)$ and $\Delta(R-I)$ color curves and the graphic representation of the model described in Section~\ref{sJ1558} at orbital phases 0.15, 0.30 and 0.80.}
\label{fJ1558}
\end{figure*}

The stellar and orbital parameters of J1558 resulting from our analysis are summarized in Table~\ref{tJ1558} and Figure~\ref{fJ1558} that details the observed and simulated light curves, color curves and the appearance of the system in different phases.

\section{J2128}\label{sJ2128}

The variability of J2128 was again discovered by \citet{norton}, who measured a period of 0.22484 days, a maximum of 14.49 mag, and a depth of 0.47 for the primary and 0.34 for the secondary minimum in the V band. \citet{lohr} report an improved period of 0.2248416 days, and that is the value used in this study. The present work is to our knowledge the first quantitative analysis of this variable.

We observed it in VRI on August 28$^{th}$, 2013 and captured two new minima given in Table~\ref{tMin}. The light curve was folded to orbital phases according to this ephemeris:

\begin{equation}\label{ephJ2128}
Min_I [HJD] = 2456533.4807(5) + 0.2248416 \times E
\end{equation}

In the analysis of the light curves of J2128 the mass ratio was again estimated with the q-search method (Figure~\ref{fQS}, bottom left), which is in this case somewhat problematic considering the relatively low orbital inclination resulting from the best-fitting model ($i\approx 73\degr$). The q-search was done for a rather large range of possible mass ratios and although the value of $q=m_2/m_1=0.40$ is the best estimate we can give with the available data, this result is to be taken with a degree of caution until it is confirmed in a radial-velocity study. The mass ratio of $q=0.40$ puts the system in a contact configuration with $f_{over}\approx 12\%$.

The color index $B-V=1.157$ corresponds to a main sequence star with the effective temperature of $T_1=4350 K$, and it is assigned to the brighter of the two components, which is in this case the larger and more massive, but cooler primary. The temperature of the secondary is calculated as a free parameter of the model and has the value of $T_2=4800 K$. Since it is the less massive star that is eclipsed in the primary minimum, we classify J2128 as a W type W~UMa binary.

Both the light curves and the color curves show significant asymmetries, which indicate the presence of spots on the stars. The optimally fitting model assumes dark spots on both components of the system, located at relatively large latitudes. The larger spot, with the temperature contrast of $T_{spot}/T_2=0.85$ is located in the polar region of the secondary star and has the greater contribution to the light curve asymmetry. The spot on the other component is of similar temperature contrast but of smaller size. Even with these spots, the fit of the model to the observations turns out to be significantly better when the exponents of gravity darkening are adjusted as free parameters. This results with the value of $\beta_2=0.17$ for the secondary component, while the optimal exponent for the primary corresponds to the expected theoretical value of $\beta_1=0.08$. The higher value of the gravity darkening exponent for the secondary star can be interpreted as a consequence of the presence of the large dark spot.

\begin{figure*}
\includegraphics{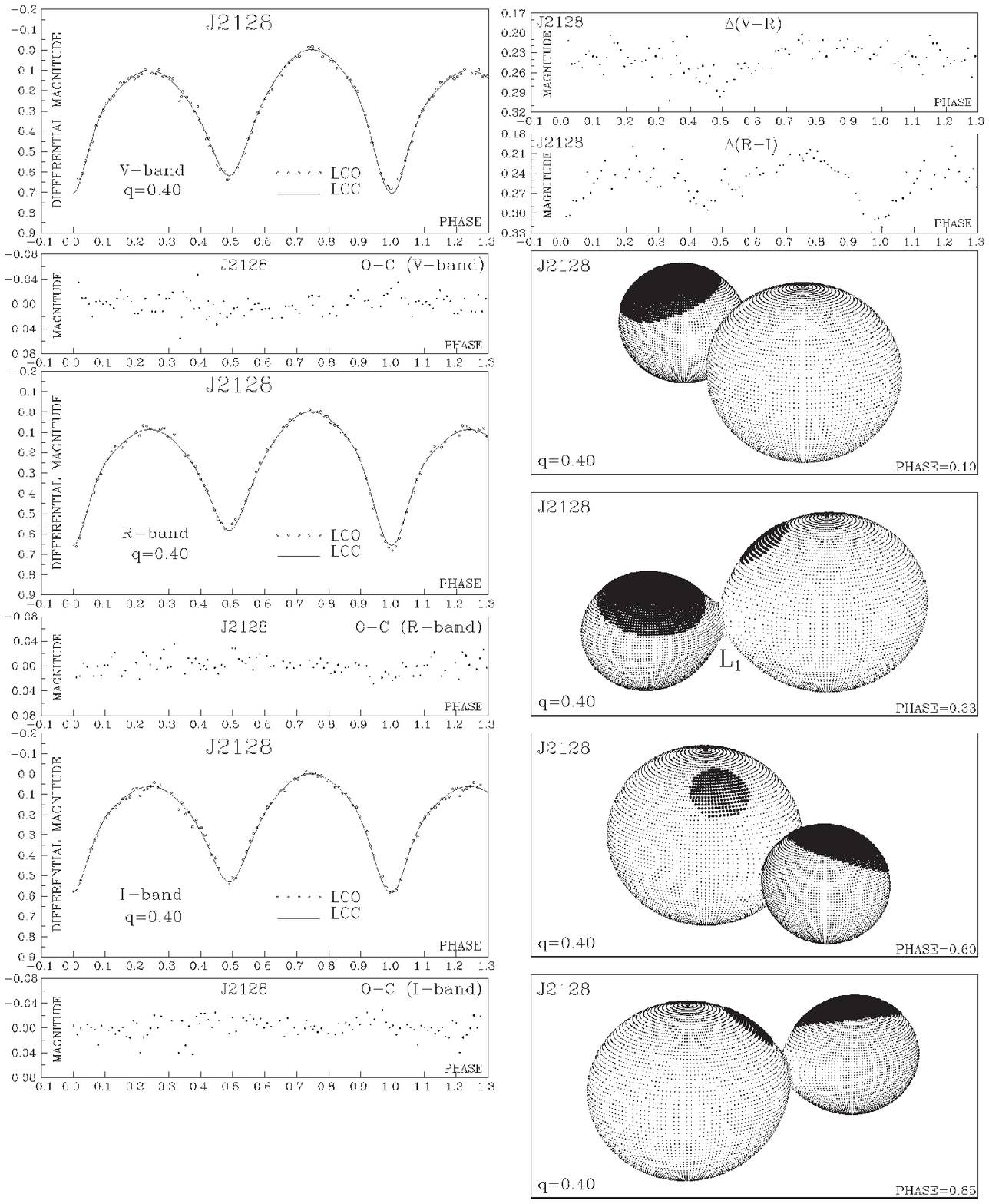}
\caption{Observed (LCO) and synthetic (LCC) light curves and the final $O-C$ residuals of {\rm J2128}, with $\Delta(V-R)$ and $\Delta(R-I)$ color curves and the graphic representation of the model described in Section~\ref{sJ2128} at orbital phases 0.10, 0.33, 0.60 and 0.85.}
\label{fJ2128}
\end{figure*}

These findings are summarized in Table~\ref{tJ2128} and Figure~\ref{fJ2128}, where the observed and simulated light curves are shown together with the color curves and the representation of the system in several orbital phases.

\section{UCAC4}\label{sUCAC4}

UCAC4 is a newly discovered binary star. Its variability was noticed and reported by one of us \citep{essam14} in the field of view for observation of J1558. The orbital period is 0.361456 days, the maximum is 15.74 mag and the light curve amplitude is 0.71 mag in the V band. Dedicated observations were conducted on three nights (May 22$^{nd}$, and June 17$^{th}$ and 18$^{th}$, 2014). The four recorded times of minimum light are given in Table~\ref{tMin}, and this is the ephemeris used to fold the light curve to phases:

\begin{equation}\label{ephUCAC4}
Min_I [HJD] = 2456827.4687(2) + 0.361456 \times E
\end{equation}

Like in the previous cases, the mass ratio of UCAC4 was determined by the q-search (Figure~\ref{fQS}, bottom right). The value of $q=m_2/m_1=0.40$ that we obtained can be considered reliable because the system has a high orbital inclination ($i\approx 87\degr$) and shows total eclipses. Light curve analysis indicates this system is in a shallow contact configuration with $f_{over}\approx 8\%$. 

The more massive star is the one eclipsed in the primary minimum, making UCAC4 an A type W~UMa binary. This is at odds with the late spectral types and the low total mass, but as in the case of J0645, we assigned the zero of orbital phase to the deeper eclipse and adopted the model that best fit the light curves.

The temperature of the primary component was estimated to be $T_1=4590$ K from the B-V color index, and the optimal model gave $T_2=4580$ K for the temperature of the secondary. Adjusting the albedos as free model parameters significantly improved the fit to the observations, resulting in smaller albedo values than expected from theory. Similarly as in the case of J1558, this was taken to indicate the existence of one or more cool spots in the neck region of the binary. After fixing the albedos to their theoretical values and adding two cool spots in the neck region, model optimization converged to a solution in which both spots are of relatively large dimensions but with a small temperature contrast ($T_{spot}/T_{1,2}=0.94$). Without these active regions we could not find an optimal fit of the model to the observations. Such anomalous temperature distribution in the neck region can again be interpreted in terms of exchange of thermal energy between stars that regardless of very different masses ($q=0.4$) have nearly the same temperatures ($\Delta T\approx 10K$).

\begin{figure*}
\includegraphics{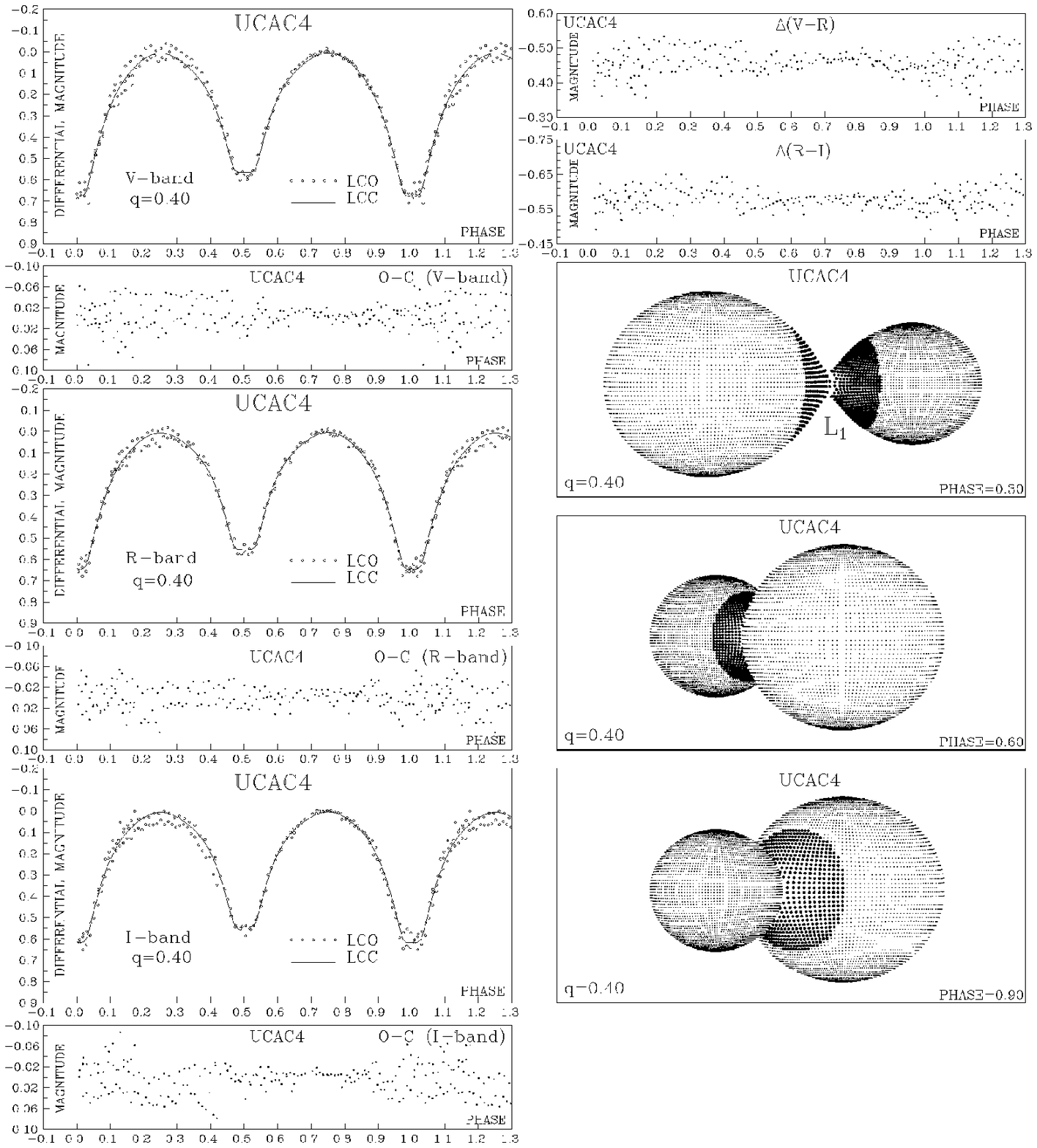}
\caption{Observed (LCO) and synthetic (LCC) light curves and the final $O-C$ residuals of {\rm UCAC4}, with $\Delta(V-R)$ and $\Delta(R-I)$ color curves and the graphic representation of the model described in Section~ \ref{sUCAC4} at orbital phases 0.30, 0.60 and 0.90.}
\label{fUCAC4}
\end{figure*}

Our results are summarized in Table~\ref{tUCAC4} and in Figure~\ref{fUCAC4}, where the observed and modeled light curves, the color curves and the appearance of the model in several orbital phases can be inspected.

\section{Comparison with other W~UMa binaries}\label{comparison}

To put these new variables in context of other W~UMa stars, we used data from two relatively recent catalogs of well-studied eclipsing binaries, combined with the findings made within our own group over the past decade for comparison:

From the catalog of \citet{A&H}, we took the spectroscopic mass ratios, orbital periods and separations, as well as absolute masses, radii, temperatures and luminosities of the components for 75 W~UMa stars. From the catalog of \citet{D&S}, we took the same quantities for 48 W~UMa stars, of which 25 are new (not contained in the first catalog). In both catalogs, the B-type variables were omitted. To this sample we added 17 W~UMa stars\footnote{
SW Lac \citep{albayrak2004, albayrak2005};
V776 Cas \citep{djur2004};
V351 Peg \citep{djur2004a};
EE Cet, XY Leo and AQ Psc \citep{djur2006};
V376 And \citep{djur2008};
V523 Cas \citep{latkovic2009};
QX And, RW Com and BD +07 3142 \citep{djur2011};
QW Gem \citep{cseki2013};
VW LMi \citep{djur2013};
AK Her, HI Dra, V1128 Tau and V2612 Oph \citep{caliskan2014}.}
studied by our group during the last decade, of which 6 are new (not contained in any of the two catalogs).

\begin{figure*}
\includegraphics{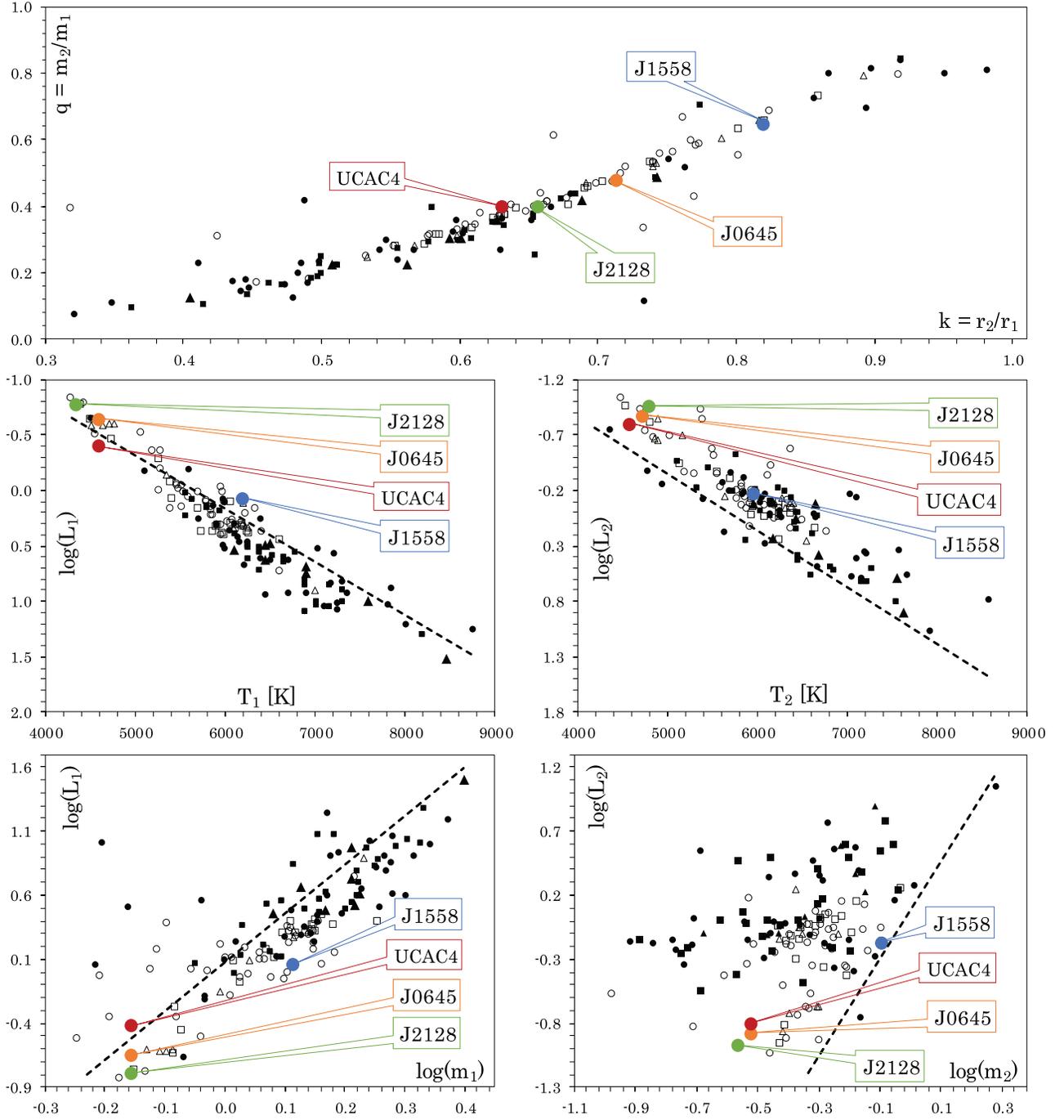}
\caption{Comparison of our targets with other W~UMa stars. Top: radius ratio vs mass ratio. Middle and bottom: HR diagram and mass-luminosity relation for the primary (left) and secondary (right) star; dashed lines represent the zero-age main sequence calibrations. Filled symbols are A, and contour symbols W type stars; circles are for data from \citet{A&H}, squares for \citet{D&S}, and triangles for past research of our group. For more details see Section~\ref{comparison}.}
\label{fComp}
\end{figure*}

All the stars in this sample have mass ratios, separations and spectral types determined from spectroscopic observations; and all of them were "solved" for orbital and stellar parameters using modeling techniques based on the Roche model. In this sense, the sample is homogeneous, and since we only wanted to see how close or far the parameter derivations for J0645, J1558, J2128 and UCAC4 are from the bulk of other W~UMa stars in the literature, no attempt was made at statistical analysis. We plotted several important correlations between stellar parameters from the sample, and examined the positions of our four stars in those plots, as shown in Figure~\ref{fComp}:

In the top panel, the spectroscopic mass ratio $q=m_2/m_1$ (where $m_1$ is always the more massive star) is plotted against the ratio of radii, $k=r_2/r_1$. The four stars from this study are plotted using photometric mass ratios that we determined. The nearly linear relation between mass ratios and size ratios that can be expected in contact binaries \citep{terrell2005} is clearly visible and our variables do not deviate from it.

In the middle and bottom panels, the mass-luminosity and temperature-luminosity relations are plotted for the primary and secondary components. The zero-age main sequence calibrations taken from \citet{allen} are shown for reference. Here too our stars do not deviate far from the catalog sample. J1558 seems to be the youngest and least evolved among our targets, but a serious discussion of ages and evolutionary states must be left for some future investigation when more data become available.

The segregation between the two subtypes of WUMa stars is evident to varying degrees in different plots of Figure~\ref{fComp}. J0645 (classified as an A type here, and as a W type by \citealt{liu14}), J2128 (W type) and UCAC4 (A type) form a loose group, while J1558 (A type) is clearly set apart; this might be another indication that J0645 and UCAC4 are in fact W types. On the other hand, the regions they occupy aren't completely devoid of A types either. This issue could easily be resolved in a radial velocity study.

\section{Resume}\label{resume}

We secured high quality multicolor light curves of four recently discovered binary stars: J0645, J1558, J2128 and UCAC4, and conducted a photometric analysis using the binary star model of G. Djura{\v s}evi{\' c} to determine the orbital and stellar properties of the systems to the best accuracy possible in the absence of complementary spectroscopic data. Our results indicate that all four systems are W~UMa type binaries with active regions (dark and bright spots) that are interpreted either as signs of magnetic activity, or when found in the neck region, as a consequence of ongoing exchange of thermal energy between the components.

A comparison with a large sample of well-studied W~UMa stars from the literature shows that the derived absolute parameters of our four targets are roughly within expected ranges. Note, however, that without spectroscopic studies which are needed for a completely reliable determination of spectral types, the mass ratio and the orbital separation in an eclipsing binary, the absolute stellar parameters of our targets found in this work are to be considered preliminary. A combined spectroscopic and photometric study of these stars would be an important next step for the understanding of low temperature contact binaries and we invite researchers with access to relevant facilities to undertake the necessary spectroscopic observations.

\acknowledgments

Research presented in this paper was funded in part by the Ministry of Education, Science and Technological Development of Republic of Serbia through the project "Stellar Physics'' (No. 176004), and by the Science and Technological Development Fund, Ministry for Scientific Research of Egypt, project ID 1335. The authors acknowledge the use of the \url[http://simbad.u-strasbg.fr/simbad/] {Simbad database}, operated at the CDS, Strasbourg, France, and \url[http://adsabs.harvard.edu/]{NASA's Astrophysics Data System Bibliographic Services}. We thank the anonymous referee whose comments and suggestions led to significant improvements of this paper and gave us ideas for future research. This work is a part of the Ph.D. thesis of Mr. Mohamed Ahmed El-Sadek.

{\it Facilities:} \facility{KAO: the 1.88 m Kottamia reflector telescope in Egypt}.

\clearpage

\begin{table*}
\caption{Summary of observational aspects.
\label{tObsLog}}
\begin{small}
\begin{tabular}{c|c|c|c|c|c|c}
\tableline
\tableline
	\multirow{2}{*}{Object}					&
	\multirow{2}{*}{Star}   					&
	\multirow{2}{*}{ID}     					&
	$\rm{RA}_{J2000}$					&
	$\rm{DEC}_{J2000}$      					&
	\multirow{2}{*}{$\rm{B}_{mag}$}			&
	\multirow{2}{*}{$\rm{V}_{mag}$}			\\
									&
									&
									&
	\scriptsize{$[hh:mm:ss.ss]$}				&
	\scriptsize{$[hh:mm:ss.ss]$}				&
									&
									\\
\tableline
	\multirow{4}{*}{J0645}					&
	\multirow{2}{*}{\scriptsize{Variable}}		&
	\scriptsize{1SWASP J064501.21+342154.9}	&
	\multirow{2}{*}{06:45:01.21}				&
	\multirow{2}{*}{+34:21:54.90}			&
	\multirow{2}{*}{15.298}				&
	\multirow{2}{*}{14.217}				\\
									&
									&
	\scriptsize{UCAC4 622-035881}			&
    									&
									&
									&
									\\ \cline{2-7}
									&
	\scriptsize{Comparison}				&
	\scriptsize{UCAC4 622-035872}			&
	06:44:58.07						&
	+34:23:52.22						&
	14.685							&
	14.037							\\ \cline{2-7}
									&
	\scriptsize{Check}					&
	\scriptsize{UCAC4 623-036530}			&
	06:44:54.60						&
	+34:24:09.71						&
	14.277							&
	13.564							\\ \cline{2-7}
\tableline
	\multirow{4}{*}{J1558}					&
	\multirow{2}{*}{\scriptsize{Variable}}		&
	\scriptsize{1SWASP J155822.10-025604.8}	&
	\multirow{2}{*}{15:58:22.09}				&
	\multirow{2}{*}{-02:56:04.85}			&
	\multirow{2}{*}{14.870}				&
	\multirow{2}{*}{14.350}				\\
									&
									&
	\scriptsize{NOMAD1 0870-0363163}		&
									&
									&
									&
									\\ \cline{2-7}
									&
	\scriptsize{Comparison}				&
	\scriptsize{NOMAD1 0870-0363204}		&
	15:58:29.59						&
	-02:59:53.72						&
	15.260							&
	13.650							\\ \cline{2-7}
									&
	\scriptsize{Check}					&
	\scriptsize{NOMAD1 0870-0363115}		&
	15:58:15.55						&
	-02:58:14.59						&
	14.930							&
	14.320							\\ \cline{2-7}
\tableline
	\multirow{4}{*}{J2128}					&
	\multirow{2}{*}{\scriptsize{Variable}}		&
	\scriptsize{1SWASP J212808.86+151622.0}	&
	\multirow{2}{*}{21:28:08.86}				&
	\multirow{2}{*}{+15:16:21.94}			&
	\multirow{2}{*}{16.108}				&
	\multirow{2}{*}{14.951}				\\ 
									&
									&
	\scriptsize{UCAC4 527-142875}			&
									&
									&
									&
									\\ \cline{2-7}
									&
	\scriptsize{Comparison}				&
	\scriptsize{UCAC4 527-142902}			&
	21:28:25.86						&
	+15:20:8.19						&
	13.230							&
	12.398							\\ \cline{2-7}
									&
	\scriptsize{Check}					&
	\scriptsize{UCAC4 527-142901}			&
	21:28:24.94						&
	+15:20:8.19						&
	13.230							&
	12.398							\\ \cline{2-7}
\tableline
	\multirow{3}{*}{UCAC4}					&
	\scriptsize{Variable}					&
	\scriptsize{UCAC4 436-062932}			&
	15:58:28.09						&
	-02:57:53.22						&
	16.121							&
	15.090							\\ \cline{2-7}
									&
	\scriptsize{Comparison}				&
	\scriptsize{UCAC4 436-062934}			&
	15:58:29.59						&
	-02:59:53.63						&
	15.840							&
	13.965							\\ \cline{2-7}
									&
	\scriptsize{Check}					&
	\scriptsize{UCAC4 436-062914}			&
	15:58:05.34						&
	-02:54:40.20						&
	13.777							&
	12.358							\\ \cline{2-7}	
\tableline
\end{tabular}
\end{small}
\end{table*}

\begin{table}
\caption{New times of minimum light derived from our observations.
\label{tMin}}
\begin{footnotesize}
\begin{tabular}{lll}
\tableline\tableline
System 		& Time of minimum [HJD] & Type 			\\
\tableline
J0645 		& 2456297.5825(3)		& Primary		\\
 			& 2456298.3283(2)		& Primary		\\
			& 2456331.2694(1)		& Secondary	\\
			& 2456331.3940(2)		& Primary		\\
			& 2456332.2620(4)		& Secondary	\\
			& 2456332.3861(4)		& Primary 		\\
J1558 		& 2456800.4656(5)		& Secondary	\\
			& 2456826.3430(3)		& Primary		\\
			& 2456826.4727(9)		& Secondary	\\
			& 2456827.3834(4)		& Primary		\\
J2128		& 2456533.3633(5)		& Secondary	\\
			& 2456533.4807(5)		& Primary		\\
UCAC4		& 2456800.5410(8)		& Secondary	\\
			& 2456826.3840(2)		& Primary		\\
			& 2456827.2873(9)		& Secondary	\\
			& 2456827.4687(2)		& Primary		\\
\tableline
\end{tabular}
\end{footnotesize}
\end{table}

\begin{table}
\caption{VRI Photometry of J0645, J1558, J2128, UCAC4}
\label{tData}
\begin{footnotesize}
\begin{tabular}{llc}
\tableline\tableline
\multicolumn{3}{c}{J0645: \textit{V} band}						\\
\tableline
HJD (d)		&	Diff. Magnitude (mag)	&	Standard Deviation	\\
2456331.38520	&	0.7098				&	0.0044		\\
2456331.38690	&	0.7591				&	0.0044		\\
2456331.38850	&	0.8066				&	0.0046		\\
2456331.39010	&	0.8202				&	0.0045		\\
2456331.39170	&	0.8374				&	0.0047		\\
2456331.39340	&	0.8534				&	0.0047		\\
2456331.39500	&	0.8471				&	0.0047		\\
2456331.39670	&	0.8396				&	0.0048		\\
2456331.39830	&	0.8353				&	0.0046		\\
2456331.39990	&	0.7890				&	0.0045		\\
2456331.40160	&	0.7452				&	0.0044		\\
2456331.40320	&	0.6982				&	0.0043		\\
\tableline
\end{tabular}
\tablecomments{This table is available in its entirety in machine-readable and
Virtual Observatory (VO) forms in the online journal. A portion is shown here
for guidance regarding its form and content.}
\end{footnotesize}
\end{table}

\begin{table}
\caption{Summary of modeling results for J0645. \label{tJ0645}}
\begin{scriptsize}
\begin{tabular}{lll}
\tableline\tableline
Properties of the fit 				& 				&				\\
\tableline
Point count 					& 469			& 				\\
$\rm \sum{(O-C)^2}$				& 0.0403			& 				\\
\tableline\tableline
System parameters 				& 				& 				\\
\tableline
P							& 0.2486159		&				\\
q 							& 0.48 $\pm$ 0.05	& 				\\
$\rm i\ [\degr]$ 					& 88.8 $\pm$ 0.5 	& 				\\
$\rm a_{orb}\ [R_{\odot}]$			& 1.7 $\pm$ 0.2 		& 				\\
F 							& 1.019 $\pm$ 0.001	& 				\\
$\rm f_{over} [\%]$ 				& 15.95 			& 				\\
$\rm \Omega_{1,2}$				& 2.7913 			&		 		\\
$\rm \Omega_{in}, \Omega_{out}$		& 2.8372, 2.5494 	& 				\\
$\rm \ell_3/(\ell_1+\ell_2+\ell_3)$	& \multicolumn{2}{l}
									  	         {0.000 $\pm$ 0.002 [V],
	 								   		0.010 $\pm$ 0.002 [R],
	 								   		0.010 $\pm$ 0.003 [I]}	\\
\tableline\tableline
Stellar Parameters 				& Primary 			& Secondary 		\\
\tableline
A 							& 0.5 			& 0.5 			\\
$\rm \beta$ 					& 0.08			& 0.08			\\
$\rm T_{eff}$ [K] 					& 4590 			& 4720  $\pm$ 20 	\\
$\rm L/(L_1+L_2)$ 				& 0.619 [V],
							    0.621 [R],
							    0.625 [I]			& 0.381 [V],
											    0.379 [R],
											    0.375 [I]		 	\\
R [D=1] 						& 0.426			& 0.305			\\
$\cal M\ \rm [M_{\odot}]$ 			& 0.7 $\pm$ 0.2 		& 0.3 $\pm$ 0.1		\\
$\cal R\ \rm [R_{\odot}]$ 			& 0.76 $\pm$ 0.07 	& 0.55 $\pm$ 0.05 	\\
$\rm log_{10}(g)$ 				& 4.5 $\pm$ 0.2 		& 4.5 $\pm$ 0.3 		\\
$\rm M_{bol}$ 					& 6.4 $\pm$ 0.2 		& 7.0 $\pm$ 0.3 		\\
\tableline\tableline
Spot parameters 					& Spot 1 (Primary) 	&  				\\
\tableline
$\rm T_{spot}/T$ 					& 0.96 $\pm$ 0.01 	&  				\\
$\rm \theta\ [\degr]$ 				& 16 $\pm$ 1 		&				\\
$\rm \lambda\ [\degr]$ 			& 320 $\pm$ 10 		&				\\
$\rm \varphi\ [\degr]$ 				& 34 $\pm$ 7 		&				\\
\tableline
\end{tabular}
\end{scriptsize}
\end{table}

\begin{table}
\caption{Summary of modeling results for J1558. \label{tJ1558}}
\begin{scriptsize}
\begin{tabular}{lll}
\tableline\tableline
Properties of the fit 				& 				&				\\
\tableline
Point count 					& 466 			& 				\\
$\rm \sum{(O-C)^2}$				& 0.2101			& 				\\
\tableline\tableline
System parameters 				& 				& 				\\
\tableline
P							& 0.2600776		&				\\
q 							& 0.65 $\pm$ 0.08	& 				\\
$\rm i\ [\degr]$ 					& 78.3 $\pm$ 0.3 	& 				\\
$\rm a_{orb}\ [R_{\odot}]$ 			& 2.2 $\pm$ 0.2 		& 				\\
F 							& 1.010 $\pm$ 0.001 	& 				\\
$\rm f_{over} [\%]$ 				& 6.92 			& 				\\
$\rm \Omega_{in}, \Omega_{out}$ 		& 3.1524, 2.7763 	& 				\\
$\rm \Omega_{1,2}$				& 3.1264 			&		 		\\
\tableline\tableline
Stellar Parameters 				& Primary 			& Secondary 		\\
\tableline
A 							& 0.5 			& 0.5 			\\
$\rm \beta$ 					& 0.08 			& 0.08 			\\
$\rm T_{eff}$ [K] 					& 6200 			& 5970 $\pm$ 20 	\\
$\rm L/(L_1+L_2)$ 				& 0.637 [V],
							    0.635 [R],
							    0.632 [I]		 	& 0.363 [V],
											    0.365 [R],
											    0.368 [I]			\\
R [D=1] 						& 0.396 			& 0.325 			\\
$\cal M\ \rm [M_{\odot}]$ 			& 1.3 $\pm$ 0.2 		& 0.8 $\pm$ 0.3 		\\
$\cal R\ \rm [R_{\odot}]$ 			& 0.94 $\pm$ 0.06 	& 0.77 $\pm$ 0.05 	\\
$\rm log_{10}(g)$ 				& 4.6 $\pm$ 0.2 		& 4.6 $\pm$ 0.2 		\\
$\rm M_{bol}$ 					& 4.6 $\pm$ 0.2 		& 5.2 $\pm$ 0.2 		\\
\tableline\tableline
Spot parameters 					& Spot 1 (Primary) 	& 				\\
\tableline
$\rm T_{spot}/T$ 					& 1.04 $\pm$ 0.02	&				\\
$\rm \theta\ [\degr]$ 				& 18 $\pm$ 5		&				\\
$\rm \lambda\ [\degr]$ 			& 351 $\pm$ 6 		&				\\
$\rm \varphi\ [\degr]$ 				& 0.0 $\pm$0.4		&				\\
\tableline
\end{tabular}
\end{scriptsize}
\end{table}

\begin{table}
\caption{Summary of modeling results for J2128. \label{tJ2128}}
\begin{scriptsize}
\begin{tabular}{lll}
\tableline\tableline
Properties of the fit 				& 				& 				\\
\tableline
Point count 					& 235 			& 				\\
$\rm \sum{(O-C)^2}$				& 0.0502			& 				\\
\tableline\tableline
System parameters 				& 				& 				\\
\tableline
P							& 0.2248416		&				\\
q 							& 0.40 $\pm$ 0.04	& 				\\
$\rm i\ [\degr]$ 					& 72.7 $\pm$ 0.3 	& 				\\
$\rm a_{orb}\ [R_{\odot}]$ 			& 1.5 $\pm$ 0.2 		& 				\\
F 							& 1.012 $\pm$ 0.002 	& 				\\
$\rm f_{over} [\%]$ 				& 11.52 			& 				\\
$\rm \Omega_{in}, \Omega_{out}$ 		& 2.6781, 2.4341 	& 				\\
$\rm \Omega_{1,2}$				& 2.6500 			&		 		\\
\tableline\tableline
Stellar Parameters 				& Primary 			& Secondary 		\\
\tableline
A 							& 0.5 			& 0.5 			\\
$\rm \beta$ 					& 0.08 			& 0.17 $\pm$ 0.02	\\
$\rm T_{eff}$ [K] 					& 4350			& 4800 $\pm$ 20	\\
$\rm L/(L_1+L_2)$ 				& 0.584 [V],
							    0.592 [R],
							    0.603 [I] 			& 0.416 [V],
											    0.408 [R],
											    0.397 [I]			\\
R [D=1] 						& 0.438 			& 0.288 			\\
$\cal M\ \rm [M_{\odot}]$ 			& 0.7 $\pm$ 0.2 		& 0.26 $\pm$ 0.09 	\\
$\cal R\ \rm [R_{\odot}]$ 			& 0.70 $\pm$ 0.06 	& 0.47 $\pm$ 0.04 	\\
$\rm log_{10}(g)$ 				& 4.6 $\pm$ 0.2 		& 4.5 $\pm$ 0.2 		\\
$\rm M_{bol}$ 					& 6.8 $\pm$ 0.2 		& 7.2 $\pm$ 0.2 		\\
\tableline\tableline
Spot parameters 					& Spot 1 (Primary) 	& Spot 2 (Secondary)	\\
\tableline
$\rm T_{spot}/T$ 					& 0.84 $\pm$ 0.02 	& 0.85 $\pm$ 0.02 	\\
$\rm \theta\ [\degr]$ 				& 15.4 $\pm$ 0.8 	& 57.3 $\pm$ 0.8 	\\
$\rm \lambda\ [\degr]$ 			& 31 $\pm$ 3 		& 293 $\pm$ 2 		\\
$\rm \varphi\ [\degr]$ 				& 36 $\pm$ 7 		& 68 $\pm$ 1 		\\
\tableline
\end{tabular}
\end{scriptsize}
\end{table}

\begin{table}
\caption{Summary of modeling results for UCAC4. \label{tUCAC4}}
\begin{scriptsize}
\begin{tabular}{lll}
\tableline\tableline
Properties of the fit 				& 				&				\\
\tableline
Point count 					& 506 			& 				\\
$\rm \sum{(O-C)^2}$				& 0.3246			& 				\\
\tableline\tableline
System parameters 				& 				& 				\\
\tableline
P							&  0.361456		&				\\
q 							& 0.40 $\pm$ 0.05	& 				\\
$\rm i\ [\degr]$ 					& 87.3 $\pm$ 0.5 	& 				\\
$\rm a_{orb}\ [R_{\odot}]$ 			& 2.1 $\pm$ 0.2 		& 				\\
F 							& 1.008 $\pm$ 0.002 	& 				\\
$\rm f_{over}$ [\%] 				& 7.71 			& 				\\
$\rm \Omega_{in},\ \Omega_{out}$ 	& 2.6781, 2.4341 	& 				\\
$\rm \Omega_{1,2}$				& 2.6593			&		 		\\
\tableline\tableline
Stellar Parameters 				& Primary 			& Secondary 		\\
\tableline
A 							& 0.5 			& 0.5 			\\
$\rm \beta$ 					& 0.08 			& 0.08 			\\
$\rm T_{eff}$ [K] 					& 4590 			& 4580 $\pm$ 20 	\\
$\rm L/(L_1+L_2)$ 				& 0.699 [V],
							  0.698 [R],
							  0.698 [I]		 	& 0.301 [V],
									  		  0.302 [R],
											  0.302 [I]			\\
R [D=1] 						& 0.436  			& 0.286  			\\
$\cal M\ \rm [M_{\odot}]$ 			& 0.7 $\pm$ 0.2 		& 0.3 $\pm$ 0.1 		\\
$\cal R\ \rm [R_{\odot}]$ 			& 1.0 $\pm$ 0.1 		& 0.65 $\pm$ 0.07 	\\
$\rm log_{10}(g)$ 				& 4.3 $\pm$ 0.2 		& 4.3 $\pm$ 0.3 		\\
$\rm M_{bol}$ 					& 5.8 $\pm$ 0.3 		& 6.7 $\pm$ 0.3 		\\
\tableline\tableline
Spot parameters 					& Spot 1 (Primary) 	& Spot 2 (Secondary)	\\
\tableline
$\rm T_{spot}/T$ 					& 0.94 $\pm$ 0.01 	& 0.94 $\pm$ 0.01	\\
$\rm \theta\ [\degr]$ 				& 38 $\pm$ 2 		& 42 $\pm$ 4 		\\
$\rm \lambda\ [\degr]$ 			& 360 $\pm$ 2 		& 178 $\pm$ 3 		\\
$\rm \varphi\ [\degr]$ 				& 0.0 $\pm$ 0.2		& 0.0 $\pm$ 0.1		\\
\tableline
\end{tabular}
\end{scriptsize}
\end{table}

\end{document}